\documentclass[aps,prl,twocolumn,groupedaddress,amsmath,amssymb,showpacs]{revtex4}
\usepackage{amssymb}
\usepackage{amsmath}
\usepackage{graphicx}
\usepackage{epsfig}
\usepackage{longtable}
\usepackage{epstopdf}
\usepackage{dcolumn}
\usepackage{bm}

\begin{document}

\preprint{Contact v.4}
\title{Spin-Orbit Symmetries of Conduction Electrons in Silicon}
\author{Pengke Li} \email{pengke@ece.rochester.edu}
\author{Hanan Dery}
\affiliation{Department of Electrical and Computer Engineering, University of Rochester, Rochester, New York, 14627}

\begin{abstract}
We derive a spin-dependent Hamiltonian that captures the symmetry of the zone edge states in silicon. We present analytical expressions of the spin-dependent states and of spin relaxation due to electron-phonon interactions in the multivalley conduction band. We find excellent agreement with experimental results. Similar to the usage of the Kane Hamiltonian in direct band-gap semiconductors, the new Hamiltonian can be used to study spin properties of electrons in silicon. \end{abstract}
\maketitle

Silicon is an ideal material choice for spintronics due to its relatively long spin relaxation time and central role in semiconductor technology. These characteristics are the reason for the wide interest in recent spin injection experiments \cite{Appelbaum_Nature07,Jonker_NaturePhysics07,Dash_Nature09,Sasaki_APE11}. To date, however, modeling of basic spin properties in silicon required elaborate numerical methods \cite{Cheng_PRL10}. Notably, the availability of transparent spin-dependent theories in direct gap semiconductors have spurred the field of semiconductor spintronics \cite{Zutic_RMP04}. The importance of a lucid theory that accurately describes spin properties of conduction electrons in silicon with relatively simple means is thus clear.

In the first part of this letter we derive a Hamiltonian that captures spin properties of conduction electrons in silicon. The Hamiltonian is constructed by its invariance to the symmetry operations of the space group, $G_{32}^2$, which describes the symmetry of the X-point at the edge of the Brillouin zone \cite{Jones_Book,Cardona_Book}. In silicon the X-point is closer to the absolute conduction band minimum than all other high symmetry points. While $k$$\cdot$$p$ and tight-binding models have been available for many decades \cite{Luttinger_PR55,Kane_JPCS57,Hensel_PR65,Cardona_PR66,Dresselhaus_PR67,Bir_Pikus_Book,Ivchenko_Pikus_Book,Jancu_PRB98,Winkler_Book}, spin has heretofore been ignored since spin-orbit coupling in Si is weak \cite{Honig_PR56,Feher_PR59,Lancaster_PPSL64,Lepine_PRB70,Zarifis_PRB87} and lattice inversion symmetry causes spin degeneracy. The present work is motivated by the emergence of experimental work on spin-polarized electron transport in silicon \cite{Appelbaum_Nature07,Jonker_NaturePhysics07,Dash_Nature09,Sasaki_APE11}.


In the second part, this Hamiltonian is used to elucidate the nature of \textit{intravalley} and \textit{intervalley} spin relaxation processes in silicon due to electron-phonon interactions. Our approach unravels the underlying physics, structure and symmetries of dominant spin-flip mechanisms. These insights cannot be shown by state-of-the art numerical studies in which only the magnitude and temperature dependence are calculated \cite{Cheng_PRL10}. We derive analytical forms and selection rules of the dominant spin-flip matrix elements and explain the subtle distinction between spin and momentum scattering processes. Importantly, it is shown that spin relaxation due to \textit{intravalley} scattering is caused by coupling of the lower and upper conduction bands (whereas intravalley momentum relaxation is governed by dilation and uniaxial deformation potentials of the lower conduction band). The accepted intravalley spin-flip matrix element derived 50 years ago by Yafet does not reveal this effect; furthermore, it was derived with only approximate spin-orbit coupling parameters and inadequate wavevector components despite correctly predicting the wavevector power-law and hence its T$^{5/2}$ dependence \cite{Yafet_1963}. Significant new insights regarding the structure of \textit{intervalley} spin relaxation are also revealed by the theory and will be discussed in this letter.

\begin{figure}
\includegraphics[width=8.5cm, height=5cm] {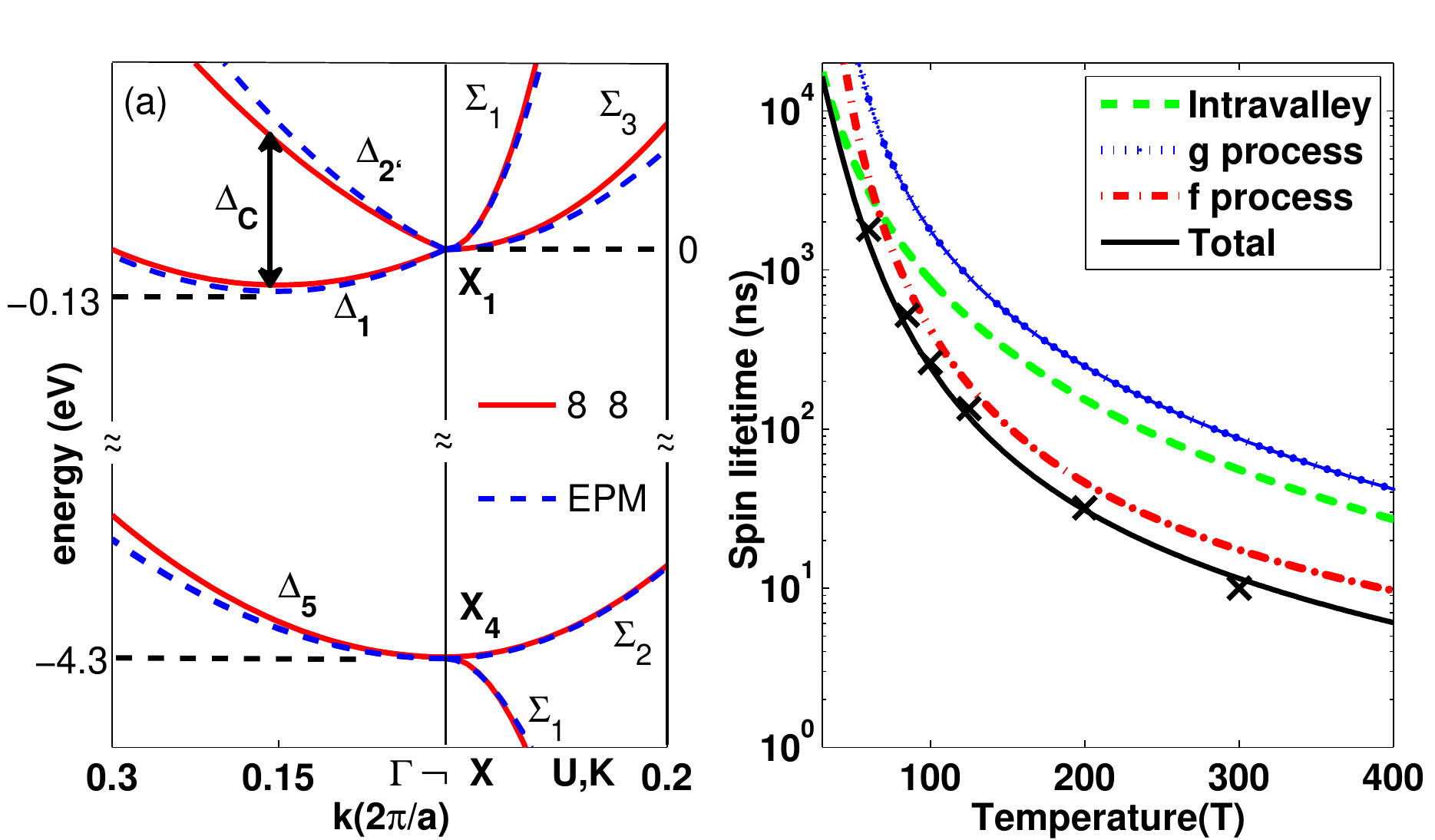}
\caption{(a) Calculated band structure near the X-point in silicon. The wavevector origin is taken at the X-point where $a$=5.43~{\AA} is the lattice constant. Solid (dash) lines are results of an empirical pseudopotential model ($8\times8$ Hamiltonian). (b) Spin relaxation in silicon due to electron-phonon interactions. Intravalley, $g$-process, and $f$-process contributions  (Eq.~(\ref{eq:ts_intra}), Eq.~(\ref{eq:ts_g}) \& Eq.~(\ref{eq:ts_f})) are denoted, respectively, by the dash green line, dotted blue line and the dash-dotted red line. The `x' symbols denote experimental results (see text).
\label{fig}}
\end{figure}

Figure~\ref{fig}(a) shows two pairs of conduction and valence bands that are pertinent to this study. Near the X-point, states and energies in this subspace are found by,
\begin{eqnarray}
H\left| \!\! \begin{array}{c}
\psi_{\mathbf{k},X_1} \\
\psi_{\mathbf{k},X_4}
\end{array}\!\! \right\rangle
=\left(
\begin{array}{cc}
H_{cc}&H_{cv}\\H_{vc}& H_{vv}
\end{array}
\right)
\left|\!\! \begin{array}{c}
\psi_{\mathbf{k},X_1} \\
\psi_{\mathbf{k},X_4}
\end{array}\!\! \right\rangle
=E
\left|\!\! \begin{array}{c}
\psi_{\mathbf{k},X_1} \\
\psi_{\mathbf{k},X_4}
\end{array} \!\! \right\rangle \,\,.
\label{eq:8 by 8}
\end{eqnarray}
Throughout this paper the crystal wavevector ($\mathbf{k}$) is taken with respect to the X-point. The upper (lower) four-components of a state, $\psi_{\mathbf{k},X_1}$ ($\psi_{\mathbf{k},X_4}$), represent coefficients of the $X_1$ ($X_4$) basis functions. These basis functions belong to the $X_1$ ($X_4$) irreducible representation of $G_{32}^2$ \cite{auxiliary}. $H_{cc}$ and $H_{\upsilon\upsilon}$ denote, respectively, conduction and valence bands contributions and $H_{c\upsilon} = H_{\upsilon c}^\dagger$ describes their coupling. These are $4\times4$ matrices due to the two-band and spin degeneracies at the X-point of diamond crystal structures \cite{footnote_timereversal}.  Since the energy gap at the X-point, $E_{g,\scriptscriptstyle{X}} \approx 4.3$~eV, is significantly larger than other energy scales, we use L\"{o}wdin partitioning \cite{Lowdin_JCM51} and lump the valence band effect on the conduction band via,
\begin{eqnarray}
\!\!\!\!\!\!\!&& \left(H_{cc}+H_{\upsilon c}^\dagger H_{\upsilon c}/E_{g,\scriptscriptstyle{X}} \right)|\psi_{\mathbf{k},X_1}\rangle_c = E_{\pm}|\psi_{\mathbf{k},X_1}\rangle_c  \label{eq:lowdin} \\
\!\!\!\!\!\!\! && |\psi_{\mathbf{k},X_4}\rangle_c = \!(E_{\pm} \!- \! H_{\upsilon\upsilon})^{\scriptscriptstyle{-1}}\!H_{\upsilon c}|\psi_{\mathbf{k},X_1}\rangle_c \! \approx \! H_{\upsilon c}/\!E_{g,\scriptscriptstyle{X}}|\psi_{\mathbf{k},X_1}\rangle_c\,, \nonumber
\end{eqnarray}
where $\pm$ refer to the upper and lower conduction bands. The spin properties of conduction electrons are set by $H_{\upsilon c}$ whereas $H_{\upsilon\upsilon}$ has a negligible effect \cite{auxiliary}. Using the method of invariants \cite{Bir_Pikus_Book} we derive $H_{cc}$ and $H_{\upsilon c}$ at the vicinity of the $X_n$-point (X-point along the $n$-axis) \cite{auxiliary},
\begin{eqnarray}
\!\!\!\!\!&H_{cc}& = \hbar^2/2m_0 ( k^2 \mathcal{I}\!\otimes\!\mathcal{I} + 2k_0k_n \mathcal{\rho}_z\!\otimes\!\mathcal{I}) \label{eq:Hcc} \\ \!\!\!\!\!&H_{\upsilon c}&=  -iP \left( k_\ell \mathcal{\rho}_y + i k_m \mathcal{\rho}_z\right)\!\otimes\!\mathcal{I} + i\Delta_{\scriptscriptstyle{X}} \left(\mathcal{\rho}_x\!\otimes\!\mathcal{\sigma}_m - \mathcal{I}\!\otimes\!\mathcal{\sigma}_\ell \right) \nonumber \\ \!\!\!\!\! &+&  \!\!\!\! \alpha\!\left[  k_n \left(i \mathcal{\rho}_z\!\otimes\!\mathcal{\sigma}_{\ell}  \!-\!  \mathcal{\rho}_y\!\otimes\!\mathcal{\sigma}_m\!\right)+\! \left(ik_\ell \mathcal{\rho}_z \!-\! k_m \mathcal{\rho}_y\right)\!\otimes\!\mathcal{\sigma}_n  \right]\!. \,\,\,\,\, \label{eq:Hvc} \end{eqnarray}
where $k_j$ denotes the $j^{th}$ component of the crystal wavevector with respect to the $X_n$-point ($\{\ell,m,n\}$ is any cyclic permutation of the $\{x,y,z\}$ crystallographic axes). $\mathcal{\sigma}_i$ and $\mathcal{\sigma}_i$$k_j$ components are due to spin-orbit coupling where $\sigma_i$ refer to Pauli matrices. $\mathcal{\rho}_i$ are invariant matrices describing the two-band degeneracy and here we choose $\mathcal{\rho}_i=\sigma_i$. $\mathcal{A}$$\otimes$$\mathcal{B}$ terms denote Kronecker products of 2$\times$2 matrices. When $\mathbf{k}$ increases toward the $\Delta$-axis, the \textit{spin-independent} basis functions of this Hamiltonian follow the compatibility relations: $X_1^1$$\rightarrow$$\Delta_1$, $X_1^{2'}$$\rightarrow$$\Delta_2'$, $X_4^{\ell}$$\rightarrow$$\Delta_5^{\ell}$ and $X_4^m$$\rightarrow$$\Delta_5^m$ (see Fig.~\ref{fig}(a) for notation and Ref.~\cite{Hensel_PR65} for details). Using this basis, the spin-independent parameters are $\hbar k_0=-\langle X_1^1| p_j| X_1^1 \rangle$=$\langle X_1^{2'}| p_j| X_1^{2'} \rangle$  and $m_0 P= \hbar | \langle X_1^{1,2'}| p_j| X_4^{j} \rangle|$ where $j=\{\ell,m\}$. In silicon the conduction band minimum position (where thermal electrons are populated) is set by $k_0$$\,$$\approx$$\,$0.15$\times$2$\pi/a$, and the mass anisotropy is set by $P$$\,$$\approx$$\,$9~eV$\cdot$${\AA}$ \cite{Hensel_PR65}.

The model spin-dependent parameters are $\Delta_{\scriptscriptstyle{X}}=\lambda | \langle X_1^{1,2'}| (\mbox{\boldmath$\nabla$}V\times\mathbf{p})_{j}|X_4^{j}\rangle |$ and $\alpha=  \hbar \lambda | \langle X_1^{1,2'}|\nabla_{j} V |X_4^{j}\rangle |$ where $\lambda=\hbar/(4m_0^2c^2)$ and $j=\{\ell,m\}$. Using an empirical pseudopotential model \cite{Chelikowsky_PRB76} we estimate $\Delta_{\scriptscriptstyle{X}}$$\sim$3.5~meV and $\alpha k_0$$\sim$1.5~meV. Substituting Eqs.~(\ref{eq:Hcc})-(\ref{eq:Hvc}) into Eq.~(\ref{eq:lowdin}), the upper and lower conduction bands state energies are,
\begin{eqnarray}
E_{\pm} \!\!= \! \frac{\hbar^2 \!k_n^2}{2m_0}\! + \! \frac{\hbar^2 \!(k_\ell^2\!+\!k_m^2)}{2m_t}\!  \pm \! \sqrt{\!E_{0,n}^2\!\!-\!E_{\ell,m}^2\!\!+\! |\eta|^2(\!k_\ell^2\!+\!k_m^2\!)}.
\label{eq:eigen_energy}
\end{eqnarray}
$m_t^{-1}=m_0^{-1}+m_{c\upsilon}^{-1}$ is the transverse mass where $m_{c\upsilon}=\hbar^2E_{g,\scriptscriptstyle{X}}/2P^2$. Other parameters are $E_{0,n}=\hbar^2k_0k_n/m_0$, $E_{\ell,m}=\hbar^2k_\ell k_m/m_{c\upsilon}$ and $\eta= 2i\Delta_{\scriptscriptstyle{X}}P/E_{g,\scriptscriptstyle{X}}$.  Eq.~(\ref{eq:eigen_energy}) includes only the leading spin-orbit term \cite{auxiliary}.

A crucial aspect of the model is that Eqs.~(\ref{eq:lowdin})-(\ref{eq:Hvc}) allow us to analytically express degenerate spin-dependent eigenstates such that $\langle \mathbf{k},\Uparrow| \sigma_z|\mathbf{k},\Downarrow\rangle$=0 \cite{Yafet_1963}. These eigenstates are represented by 8-component normalized vectors. The components are coefficients of the $X_n$-point basis functions: $\{X_1^{2'}$$\uparrow$,~$X_1^{2'}$$\downarrow$, $X_1^{1}$$\uparrow$, $X_1^{1}$$\downarrow$, $X_4^{\ell}$$\uparrow$, $X_4^{\ell}$$\downarrow$, $X_4^{m}$$\uparrow$, $X_4^{m}$$\downarrow\}$. In the $n$=$z$ case (parallel to the spin quantization axis), the vectors in the \textit{\underline{lower}} conduction band read,
\begin{widetext}
\begin{eqnarray}
\!\!\!|\mathbf{k},\Uparrow\rangle \!\! &\simeq& \!\! \left[ \frac{\sqrt{2}E_{x,y}}{\sqrt{E_C(2E_{0,z}\!\!+\!E_C)}},\, \frac{\sqrt{2}\eta(k_x-ik_y)}{\sqrt{E_C(2E_{0,z}\!\!+\!E_C)}},\, \sqrt{\frac{2E_{0,z}\!\!+\!E_C}{2E_C}},\, \frac{i\eta'(k_x\!+\!ik_y)}{2E_{g,\scriptscriptstyle{X}}},\, -\frac{P k_x}{E_{g,\scriptscriptstyle{X}}},\, -\frac{\Delta_{\scriptscriptstyle{X}}'}{E_{g,\scriptscriptstyle{X}}},\, -\frac{P k_y}{E_{g,\scriptscriptstyle{X}}},\, -\frac{i\Delta_{\scriptscriptstyle{X}}'}{E_{g,\scriptscriptstyle{X}}} \right]^T\!\!,\,\,\,\,\, \label{eq:k_up_z} \\
\!\!\!\langle\mathbf{k},\Downarrow| \!\! &\simeq& \!\! \left[ -\frac{\sqrt{2}\eta(k_x-ik_y)}{\sqrt{E_C(2E_{0,z}\!\!+\!E_C)}},\, \frac{\sqrt{2}E_{x,y}}{\sqrt{E_C(2E_{0,z}\!\!+\!E_C)}},\, \frac{-i\eta'(k_x\!+\!ik_y)}{2E_{g,\scriptscriptstyle{X}}},\, \sqrt{\frac{2E_{0,z}\!\!+\!E_C}{2E_C}},\,  \frac{\Delta_{\scriptscriptstyle{X}}'}{E_{g,\scriptscriptstyle{X}}},\, -\frac{P k_x}{E_{g,\scriptscriptstyle{X}}},\, \frac{i\Delta_{\scriptscriptstyle{X}}'}{E_{g,\scriptscriptstyle{X}}},\, -\frac{P k_y}{E_{g,\scriptscriptstyle{X}}}  \right]\!\!,\,\,\,\,\, \label{eq:k_down_z}
\end{eqnarray}
\end{widetext}
where $\Delta_{\scriptscriptstyle{X}}'$=$\Delta_{\scriptscriptstyle{X}}$+$\alpha|k_z|$, $\eta'$=$2$$i$$\Delta_{\scriptscriptstyle{X}}'$$P$/$E_{g,\scriptscriptstyle{X}}$, and $E_C$ is the energy spacing between the conduction bands (twice the square root in Eq.~(\ref{eq:eigen_energy})). For later use, we define an important parameter $\Delta_C \equiv E_C(\mathbf{k}_0) \simeq 2\hbar^2k_0^2/m_0\sim$~0.5~eV which denotes the energy spacing at the valley center (see Fig.~(\ref{fig})). State expressions in the $x$ and $y$ valleys are provided in the supplementary material \cite{auxiliary}. We can, however, unify features along all crystallographic axes by defining $\kappa$,
\begin{eqnarray}
 \kappa \equiv \left\{
  \begin{array}{l l}
    k_y,  & \quad \text{if $n$=$x$\,,}\\
    ik_x & \quad \text{if $n$=$y$}\,,\\
    k_x + ik_y & \quad \text{if $n$=$z$\,.}
  \end{array} \right.\label{eq:kappa}
\end{eqnarray}

We first briefly discuss the spin mixing which is a measure of the total magnitude of spin-down components in a $|$$\Uparrow$$\rangle$ state (and vice versa). The spin mixing can reach its maximal value (1/2) along certain directions at the edge of the Brillouin zone where it is given by,
\begin{eqnarray}
\!\!\!\!\!\!\!\!\! \beta(\!k_n\!=\!0) = \!(1\!+\!\delta_{n,z})\!\frac{\Delta_{\scriptscriptstyle{X}}^2}{E_{g,\scriptscriptstyle{X}}^2} \!+\! \frac{1}{2} \frac{ |\eta \kappa|^2}{E_{\ell,m} + |\eta|^2(k_\ell^2+k_m^2)}.
\label{eq:mix_1}
\end{eqnarray}
This result elucidates the nature of the spin hot-spot \cite{Fabian_PRL98,Cheng_PRL10}. At the vicinity of the valley center the mixing is of the order of $10^{-6}$ and it is given by,
\begin{eqnarray}
\!\!\! \beta(k_n \sim k_0) = \!(1+\delta_{n,z})\!\left(\frac{\Delta_{\scriptscriptstyle{X}}+\alpha k_0}{E_{g,\scriptscriptstyle{X}}-\tfrac{\Delta_C}{4}}\right)^2 \!\!\! + \! \left(\frac{2|\eta \kappa|}{\Delta_C}\right)^2\!\!.
\label{eq:mix}
\end{eqnarray}


A powerful application of the theory is in elucidating the spin relaxation mechanisms. We focus on intrinsic or non-degenerate \textit{n}-type silicon where spin relaxation is governed by electron-phonon interaction across a wide temperature range \cite{Cheng_PRL10,Yafet_1963,Elliott_PR54}. The relaxation rate is,
\begin{eqnarray}
\frac{1}{\tau_{s,\nu}} \!&=& \!\frac{2\pi\hbar}{\varrho N_c} \! \int \!\! d^3 \mathbf{k} e^{-E_\mathbf{k}/k_BT} \!\!\int \!\! \frac{d^3\mathbf{k}'}{(2\pi)^3} \big| M_\nu^{sf}(\mathbf{k},\mathbf{k}') \big|^2 \,\,\,\,\,\,\,\,\,  \label{eq:ts_phonon}  \\    &\times& \!\! \frac{1}{\Omega_{\nu}(\mathbf{q})} \! \Big[ \sum_{\pm}(n_{\nu,\mathbf{q}} + \tfrac{1}{2} \pm \tfrac{1}{2}) \delta( E_{\mathbf{k}'}  - E_{\mathbf{k}} \pm \Omega_{\nu}(\mathbf{q}) ) \!   \Big] \,. \nonumber
\end{eqnarray}
where $\varrho=2.33$~gr/cm$^3$ is the crystal density and $N_c=(2\pi m_d k_BT/\hbar^2)^{3/2}$ is an effective density constant ($m_d^3=m_0m_t^2$). $\nu$, $\mathbf{q}=\mathbf{k}-\mathbf{k}'$, $\Omega_{\nu}(\mathbf{q})$ and $n_{\nu,\mathbf{q}}$  denote the phonon mode, wavevector, energy and Bose-Einstein distribution, respectively.  The $+$($-$) refers to phonon emission (absorption) processes. $M_\nu^{sf}(\mathbf{k},\mathbf{k}')=\langle \mathbf{k}',\Downarrow| M_{\nu,\mathbf{q}} |\mathbf{k},\Uparrow\rangle$  is the spin-flip matrix element between electronic states in the lower conduction band. A central point in spin relaxation of silicon is that mechanisms that dominate the momentum relaxation are not identical to those that lead to spin relaxation. In the supplementary information the Hamiltonian model is used to derive selection rules of various spin relaxation processes \cite{auxiliary}. Here we derive explicit electron-phonon interaction forms and their ensuing spin relaxation times.

\textit{Intravalley scattering.} Inspection of the electronic states reveals that the intraband spin-flip coupling is much smaller than the interband coupling between conduction bands. For example, at the valley center region ($2E_{0,z} \rightarrow \!E_C$ in Eqs.~(\ref{eq:k_up_z})-(\ref{eq:k_down_z})) the product square amplitude of the dominant $X_1^1$$\uparrow$ coefficient in $|$$\mathbf{k}$,$\Uparrow$$\rangle$ with the $X_1^1$$\uparrow$ coefficient in $\langle$$\mathbf{k}'$,$\Downarrow$$|$ is smaller than with the $X_1^{2'}$$\uparrow$ coefficient in $\langle$$\mathbf{k}'$,$\Downarrow$$|$ by a factor of $(\Delta_C/E_{g,\scriptscriptstyle{X}})^2 \simeq 1/64$. Interband coupling between $X_1^1$ and $X_1^{2'}$ states is feasible via the deformation potential, $\Xi_i$, associated with the off-diagonal strain component ($e_{\ell m}$ where $n$ is the valley axis) \cite{Hensel_PR65,footnote_interbanddformation}. The symmetries of this coupling result in a dominant role of the transverse acoustic (TA) phonon mode \cite{auxiliary} and the spin-flip matrix element reads,
\begin{eqnarray}
|M_{TA}^{sf}(\mathbf{k},\mathbf{k}')|_{n,n} = \frac{|\sqrt{2}\eta(\kappa-\kappa')|}{\Delta_c} |\mathbf{k}-\mathbf{k}'| \Xi_i     \,\,. \label{eq:intra_mat}
\end{eqnarray}
Intravalley momentum scattering, on the other hand, are governed by the intraband dilation and uniaxial deformation potentials ($\Xi_d$ \& $\Xi_u$) that lead to a dominant role of longitudinal acoustic (LA) phonons \cite{Herring_PR56}. In Eq.~(\ref{eq:intra_mat}) we have used a single effective deformation potential $\Xi_i\simeq 8$~eV \cite{Laude_PRB71} which absorbs the effect of the valleys' ellipsoidal energy dispersion. Substituting Eq.~(\ref{eq:intra_mat}) into Eq.~(\ref{eq:ts_phonon}) and using the long wavelength limit $\Omega_{\scriptscriptstyle{TA}}(\mathbf{q})=\hbar \upsilon_{\scriptscriptstyle{TA}}\mathbf{q} \ll k_BT$ where $\upsilon_{TA} \simeq 5\cdot10^5$~cm/sec is the TA phonon speed, we get the average intravalley spin relaxation rate,
\begin{eqnarray}
\frac{1}{\tau_{s,i}} \!=\! \frac{128}{9} \frac{m_t}{m_{c\upsilon}} \!\left(\!\frac{\Delta_{\scriptscriptstyle{X}}}{\Delta_C}\!\right)^{\!\!2}   \!\! \left(\frac{2 m_d}{\pi}\right)^{\!\frac{3}{2}} \frac{ \Xi_i^2 (k_BT)^{\frac{5}{2}}}{  \hbar^4 \varrho \upsilon_{\scriptscriptstyle{TA}}^2  E_{g,\scriptscriptstyle{X}} } \,\,. \,\,
\label{eq:ts_intra}
\end{eqnarray}
The $T^{\frac{5}{2}}$ dependence was predicted by Yafet (see pp.~75-80 in Ref.~\cite{Yafet_1963}). However, our theory reveals the correct magnitude and hidden symmetries: coupling between conduction bands, TA mode dominant role, and the involved wavevector components as shown in Eq.~(\ref{eq:intra_mat}) via the $\kappa$ parameter (defined in Eq.~(\ref{eq:kappa})) rather than $|$$\mathbf{k}$-$\mathbf{k}'$$|^2$.

\textit{Intervalley $g$-process scattering.} In this umklapp process a phonon mode with wavevector $\mathbf{q}_g$$\sim$0.3$\times($2$\pi/a$)$\hat{n}$ is needed to scatter electrons between the ($\pm$$k_0$)$\hat {n}$ valleys \cite{Cardona_Book}.  $g$-process momentum scattering is dominated by interaction with longitudinal optical phonons \cite{Streitwolf_pss70}. The dominant $g$-process spin-flips, on the other hand, are governed by interaction with acoustic phonons which are forbidden at zero-order for momentum scattering \cite{auxiliary}. To understand this behavior, we note that replacing $k_n\rightarrow -k_n$ in the Hamiltonian of the $-n$ valley (Eqs.~(\ref{eq:Hcc})-(\ref{eq:Hvc})) leads to an exchange of coefficients between $X_1^1 \leftrightarrow X_1^{2'}$ and  $X_4^\ell \leftrightarrow X_4^m$ states in Eqs.~(\ref{eq:k_up_z})-(\ref{eq:k_down_z}). As a result, the dominant spin-flip mechanism during a $g$-process is governed by intraband coupling albeit at opposite valleys (i.e., between respective $X_1^1$ coefficients of $|$$\mathbf{k}$,$\Uparrow$$\rangle$ in the $n$~valley and $\langle$$\mathbf{k}'$,$\Downarrow$$|$ in the $-n$~valley). The resulting spin-flip matrix element is,
\begin{eqnarray}
|M_{g}^{sf}(\mathbf{k},\mathbf{k}')|_{n,-n} =   D_g  \frac{ |\sqrt{2}\eta(\kappa+\kappa')|}{\Delta_c}     \,\,, \label{eq:g_mat}
\end{eqnarray}
Note that when $\kappa = -\kappa'$ the matrix element is zero in accord with time reversal symmetry. The large longitudinal component $|\mathbf{k}-\mathbf{k}'|_n \approx 2k_0$ relates to the dilation and uniaxial deformation potential constants via $D_g \approx 2k_0(\Xi_d + \Xi_u) \approx 4$~eV/$\AA$. As mentioned this coupling is associated with LA phonon modes where $\Omega_g = \Omega_{LA}(\mathbf{q}=2k_0\hat{n}) \approx$~21~meV. Substituting Eq.~(\ref{eq:g_mat}) into Eq.~(\ref{eq:ts_phonon}) we get the average $g$-process spin relaxation rate,
\begin{eqnarray}
\frac{1}{\tau_{s,g}} = \frac{64}{9} \frac{m_t}{m_{c\upsilon}} \! \left(\!\frac{\Delta_{\scriptscriptstyle{X}}}{\Delta_C}\!\right)^{\!\!2} \!\! \left(\frac{2m_d}{\pi}\right)^{\!\!\frac{3}{2}} \!  \frac{\sqrt{\Omega_g} D_g^2}{\hbar^2 \varrho E_{g,\scriptscriptstyle{X}} } \cdot \frac{g(y)}{\exp(y)-1} ,\,\,\,\
\label{eq:ts_g}
\end{eqnarray}
where $y=\Omega_g/k_BT$ and $g(y)$ is a sum of two confluent hypergeometric functions of the second kind,
\begin{eqnarray}
g(y) \!=\! \tfrac{\sqrt{\pi y^3}}{2}\!\left(U\!\!\left(\tfrac{3}{2},3;y\right)+ 3U\!\!\left(\tfrac{5}{2},4;y\right)\!\right)\!\approx \!1\!+\!5y^{\!-\frac{3}{2}}.
\end{eqnarray}

\textit{Intervalley $f$-process scattering.} In this umklapp process a phonon mode with wavevector $\mathbf{q}_f \approx k_0\hat{n}+k_0\hat{\ell}+(2\pi/a)\hat{m}$ is needed to scatter electrons between valleys that reside in the $\hat{n}$ and $\hat{\ell}$ directions \cite{Cardona_Book}. This wavevector resides on the $\Sigma$ axis where spin (momentum) scattering is dominated by phonon modes with $\Sigma_{1}$ \& $\Sigma_{3}$ ($\Sigma_1$) symmetries \cite{auxiliary}. Spin relaxation is unique since it is carried via coupling of valence and conduction bands. This coupling is a result of the non-orthogonal bases of the $n$ and $\ell$ valleys that are involved in the transition (i.e., $\langle X_{1,n}^1|X_{4,\ell}^m \rangle \neq 0$). The spin-flip matrix element reads,
\begin{eqnarray}
|M_{f}^{sf}(\mathbf{k},\mathbf{k}')|_{\ell,n} =  C_i D_i     \frac{ \Delta_{\scriptscriptstyle{X}} +\alpha k_0}{E_{g,\scriptscriptstyle{X}}}    \,\,. \label{eq:f_mat}
\end{eqnarray}
$C_1$=2 and $C_3$=1 if one of the involved valleys ($n$ or $\ell$) is collinear with the spin quantization axis ($z$) or $C_1$=0 and $C_3$=$\sqrt{2}$ if both valleys are perpendicular to it \cite{auxiliary}. $D_i$ is a scattering constant associated with a phonon mode of $\Sigma_i$ symmetry in an $f$-process. Using an empirical pseudopotential model, an adiabatic bond-charge model and a rigid-ion approximation (similar to the procedure in Ref.~\cite{Cheng_PRL10}), the calculated values are $D_1 \approx 12$~eV/$\AA$ and $D_3 \approx 5$~eV/$\AA$. The $\Sigma_{1(3)}$ symmetry is governed by the upper (middle) acoustic branch with a phonon energy of $\Omega_{f,1}\approx$~47~meV ($\Omega_{f,3}\approx$~23~meV). Using Eq.~(\ref{eq:f_mat}) and the possibility of scattering to 4 valleys we get the average $f$-process spin relaxation rate,
\begin{eqnarray}
\frac{1}{\tau_{s,f}} \!= \!   \frac{16}{3}\!\! \left(\!\!\frac{\Delta_{\!\scriptscriptstyle{X}}'\!(k_0)}{E_{g,\scriptscriptstyle{X}}}\! \right)^{\!\!2} \! \! \! \left(\!\frac{2m_d}{\pi}\!\right)^{\!\!\frac{3}{2}} \!\!\!  \sum\limits_{i=1,3}  \frac{ A_i D_i^2}{\hbar^2 \varrho \sqrt{\Omega_{f,i}}} \frac{f(y_i)}{\exp(y_i)\!-\!\!1},\,\, \label{eq:ts_f}
\end{eqnarray}
where $\Delta_{\scriptscriptstyle{X}}'(k_0)$$\,$=$\,$$\Delta_{\scriptscriptstyle{X}}$$\,$+$\,$$\alpha k_0$, $A_1$=2, $A_3$=1, $y_i$=$\Omega_{f,i}/k_BT$. $f(y)$$\,$=$\,$$\sqrt{y}\,exp(y/2)K_{-1}(y/2)$ is associated with the modified Bessel function of the second kind ($\sqrt{\pi}$$\,$$\lesssim$$\,$$f(y_i)$$\,$$\lesssim$$\,$3 when 10$\,$K$\,$$<$$\,$T$\,$$<$$\,$400$\,$K). Figure~\ref{fig}(b) shows the spin relaxation of all mentioned processes as a function of temperature. Spin relaxation is dominated by intravalley ($f$-process) scattering in low (high) temperatures. The figure shows excellent agreement with experimental results which have used electron spin resonance (`x' symbols  at T$>$150~K) \cite{Lancaster_PPSL64,Lepine_PRB70}, and spin-transport via a Larmor-clock analysis \cite{Jang_PRL09} and spin-valve magnetoresistance \cite{private}.

In conclusion, we have derived a Hamiltonian that elucidates the spin properties of conduction electrons in silicon. Applications of the Hamiltonian were used to extract analytical spin relaxation times and to explain the electron-phonon mechanisms that dictate the relaxation. The theory also establishes a solid ground to analytically study spin relaxation in doped silicon via scattering with impurities or via exchange with holes. In addition, straightforward extensions can be made to describe stressed silicon (incorporating strain invariant parameters) or to study spin properties in silicon heterostructures and nanostructures by plane wave expansions. Finally, the theory guide experimental studies of spin properties by providing lucid insights into the various scattering mechanisms. New experiments can be designed to extract the spin-orbit coupling parameters ($\Delta_X$ and $\alpha$).

This work is supported by AFOSR Contract No. FA9550-09-1-0493 and by NSF Contract No. ECCS-0824075. We deeply thank Mr. Y. Song for providing elaborate numerical results which served as a crucial reference for our analytical results.


\begin{thebibliography}{99}
\bibitem{footnote_interbanddformation} Intravalley spin-flips via coupling of $X_1$ and $X_4$ coefficients are negligible due to a much smaller conduction to valence bands deformation potential ($\Xi_i^2 \gg \Xi_{cv}^2$).
\end{thebibliography}
\end{document}